# CONVEX PRICING BY A GENERALIZED ENTROPY PENALTY[1]

By Johannes Leitner

*Vienna University of Technology*

In an incomplete Brownian-motion market setting, we propose a convex monotonic pricing functional for nonattainable bounded contingent claims which is compatible with prices for attainable claims. The pricing functional is defined as the convex conjugate of a generalized entropy penalty functional and an interpretation in terms of tracking with instantaneously vanishing risk can be given.

**1. Introduction.** Given an incomplete market, one of the problems of mathematical finance is to price nonattainable contingent claims. One way to do this is (expected or robust) utility indifference pricing. Typically, pricing functionals are desired to be convex, monotonic, (weakly) continuous and translation invariant or monetary. Furthermore, for attainable claims, the pricing functional should lead to the price of a replicating self-financing hedging strategy.

However, having sold a nonattainable contingent claim for such a utility indifference price, it is not clear whether their exists a good way to hedge the claim from a (market) risk management point of view.

In our approach, the risk stemming from not being able to perfectly replicate a nonattainable claim, measured at an instantaneous level, directly enters the pricing functional via an instantaneous penalty. The pricing functional can be represented using its convex conjugate which can be interpreted as a generalized relative entropy functional. The total penalty turns out to be the generalized entropy of an equivalent martingale measure depending on the claim, relative to the minimal martingale measure, introduced in Föllmer and Schweizer (1990).

Similarly, as for expected exponential utility indifference pricing functionals [see Rouge and El Karoui (2000), Lazrak and Quenez (2003) and

Received August 2006; revised August 2007.

[1]Supported by the Christian Doppler Research Association (CDG).

Mania and Schweizer (2005)], a quadratic BSDE appears. See El Karoui, Peng and Quenez (1997), El Karoui and Mazliak (1997), Ma and Yong (1999) and Peng (2004) for the BSDE theory. Quadratic BSDEs have been considered in Kobylanski (1997), Kobylanski (2000), Lepeltier and San Martin (1998) and Briand and Hu (2006).

After introducing the market model in Section 2, we propose in Section 3 a generalized relative entropy functional. This convex functional is used in Section 4 as a penalty functional in order to define a convex pricing functional as its convex conjugate. Alternatively, the pricing functional can be defined as the initial value of a bounded quadratic growth BSDE with terminal value in $L^\infty$. In Section 5 a local representation of the pricing functional is shown to also make it an instantaneous risk measure.

**2. Preliminaries.** Let $\mathbf{\Omega} := (\Omega, \mathcal{F}, \mathbf{F}, P)$ be a stochastic base satisfying the usual assumptions, where the augmented filtration $\mathbf{F} = (\mathcal{F}_t)_{t \in \mathbf{R}_+}$ is generated by an $\mathbf{R}^n$-dimensional standard Brownian motion $W = (W^1, W^2)$, $W^k$ $\mathbf{R}^{n_k}$-valued, $n_k \geq 1, k = 1, 2$, $\langle W^1, W^2 \rangle = 0$. In particular, we assume $(\Omega, \mathcal{F}_0, P)$ to be a complete probability space, an assumption which will be needed for several measurability results for closed random sets.

Consider a market model with $n_1$ traded assets whose price processes are, for sake of simplicity, modeled as $S = S_0 + \lambda \cdot \mathbf{t} + W^1$ for a uniformly bounded predictable $\mathbf{R}^{n_1}$-valued process $\lambda$, $S_0 \in \mathbf{R}^{n_1}$, and where $\mathbf{t}_t := t, t \in \mathbf{R}_+$. For a general price process with martingale part given as a stochastic integral with respect to $W^1$, and under a regularity assumption on its drift and volatility matrix, by orthogonalization, this can always be achieved.

The *relative entropy* of a probability measure $Q$ with respect to $P$ is defined for $Q \ll P$ as

$$(1) \qquad H(Q|P) := E_P\left[\frac{dQ}{dP} \ln \frac{dQ}{dP}\right] = E_Q\left[\ln \frac{dQ}{dP}\right] \in [0, \infty],$$

respectively as $H(Q|P) := \infty$ otherwise; see, for example, Csiszár (1975).

Fix a time horizon $0 < T < \infty$ and denote by $\mathcal{Q}$ the set of all probability measures on $(\Omega, \mathcal{F}_T)$, absolutely continuous with respect to $P$. Define $\mathcal{M}$ as the set of probabilities $Q \in \mathcal{Q}$ such that $S$ becomes on $[0, T]$ a local martingale with respect to $Q$. The so-called minimal equivalent martingale measure (EMM) $Q^{\min}$ over $[0, T]$ is defined by $\frac{dQ^{\min}}{dP} = \mathcal{E}(-\lambda \cdot W^1)_T$; see Föllmer and Schweizer (1990). $Q^{\min}$ is in general different from the minimal entropy EMM; see Frittelli (2000) and Delbaen et al. (2002).

Since $Q^{\min} \sim P$, we have $Q \ll Q^{\min}$ for all $Q \in \mathcal{Q}$ and we can interpret $\mathcal{Q}$ as a subspace of $L^1(Q^{\min})$ via $Q \mapsto \frac{dQ}{dQ^{\min}}$. Note that $\mathcal{M}$ is convex and closed in $L^1(Q^{\min})$. Recall that, by the predictable martingale representation property of Brownian motion, we can find for all $Q \in \mathcal{M}$ with density process $q := q^Q := E[\frac{dQ}{dP}|\mathcal{F}.]$, an up to evanescence unique predictable $\mathbf{R}^{n_2}$-valued process $\gamma = \gamma^Q$ with $\gamma = \gamma \mathbf{1}_{[0,\tau)}$, where $\tau := \tau^Q := \inf\{t \geq 0 | q_t = 0\}$, such that $\gamma \cdot W^2$ is a local martingale on $[0, \tau)$ and $q = \mathcal{E}(-\lambda \cdot W^1 + \gamma \cdot W^2)$.

**3. Generalized entropy penalties.** Set $h^E := \frac{|\gamma|^2}{2} \cdot \mathbf{t}$. $h^E$ is the entropy-Hellinger process of $Q \in \mathcal{M}$ with respect to $Q^{\min}$ introduced in Choulli and Stricker (2005). It has been shown there that $H(Q|Q^{\min}) = E_Q[h_T^E]$ holds. The idea is now to replace $\frac{dh^E}{dt} = \frac{|\gamma|^2}{2}$ by a more general convex function $\hat{\rho} \geq 0$ of $\gamma$ (to be specified more precisely in the next section), and to define a generalized entropy of $Q$ with respect to $Q^{\min}$ by

$$
\begin{aligned}
H^\rho(Q|Q^{\min}) &:= E_Q\bigg[\int_0^T \hat{\rho}_t(\gamma)\,dt\bigg] \\
&= E\bigg[\int_0^T q_t \hat{\rho}_t(\gamma)\,dt\bigg] \in [0, \infty],
\end{aligned}
\tag{2}
$$

where $\hat{\rho}$ denotes the convex conjugate of $\rho$ and where we have used the following well-known result for the 2nd identity:

LEMMA 3.1. *Let $A$ be a nondecreasing predictable (continuous) process with $A_0 = 0$. Then for all $Q \ll P$ with density process $q$, we have $E_Q[A_\infty] = E[\int_0^\infty q_t\,dA_t] \in [0, \infty]$.*

PROOF. Note that $E_Q[A_{T_n}] \to E_Q[A_\infty]$ for all increasing sequences of stopping times $(T_n)_{n \geq 1}$ with $\lim_{n \to \infty} T_n = \infty$. Since $qA = A \cdot q + q \cdot A$ and $A \cdot q$ is a local martingale, we find for a localizing sequence $(T_n)_{n \geq 1}$, $E_Q[A_{T_n}] = E[q_{T_n} A_{T_n}] = E[q \cdot A_{T_n}] \to E[q \cdot A_\infty]$. □

REMARK 3.1. It is possible to show convexity of the functional $H^\rho(\cdot|Q^{\min})$ directly. However, establishing weak lower semi-continuity by a direct argument seems to be difficult (due to the complex relationship between $Q$ and $\gamma$, especially if $Q \in \mathcal{Q}$ is not equivalent to $P$). Only by identifying $H^\rho(\cdot|Q^{\min})$ as the convex conjugate of a weak-∗ lower semi-continuous convex functional are we going to achieve this in Theorem 4.1.

3.1. *Closed random sets.* Let us collect some well-known results on random closed sets and normal integrands; see Molchanov (2005), in particular, Chapter 5.3.1.

Set $\tilde{\Omega} := \Omega \times \mathbf{R}_+$, denote by $\mathcal{P}$ the predictable $\sigma$-algebra for $\boldsymbol{\Omega}$ on $\tilde{\Omega}$ and let $\tilde{P}$ denote the product measure of $P$ and the Lebesgue-measure on $\mathbf{R}_+$. Let $\rho$ be a $\mathcal{P}$-measurable *normal convex integrand* on $\mathbf{R}^{n_2}$, that is, $\rho \colon \tilde{\Omega} \times \mathbf{R}^{n_2} \to \mathbf{R}$, and $\tilde{\omega} \mapsto \text{epi}(\rho(\tilde{\omega}, \cdot)), \tilde{\omega} \in \tilde{\Omega}$, is a $\mathcal{P}$-measurable random closed convex set, where $\mathrm{epi}(\rho(\tilde{\omega},\cdot)) := \{(z,r) \in \mathbf{R}^{n_2} \times \mathbf{R} | \rho(\tilde{\omega},z) \leq r\}$ for all $\tilde{\omega} \in \tilde{\Omega}$. For a $\mathbf{R}^{n_2}$-valued predictable process $\gamma$, $\rho(\gamma):\tilde{\Omega} \to \mathbf{R}$ defined by $\tilde{\omega} \mapsto \rho(\tilde{\omega},\gamma(\tilde{\omega}))$ is then $\mathcal{P}$-measurable. Similarly, as in Theorem 5.3.13 in Molchanov (2005), one shows that $\partial\rho(\gamma)$, where the closed convex random set $\partial\rho(z) := \{y \in \mathbf{R}^{n_2} | \rho(z) + \langle x,y \rangle \leq \rho(z+x), \forall x \in \mathbf{R}^{n_2}\}$ is the (random) sub-differential of $\rho$ at $z \in \mathbf{R}^{n_2}$, is a $\mathcal{P}$-measurable closed convex random set. Note that since $\rho$ is assumed to be $\mathbf{R}$-valued, it follows that $\tilde{P}$-a.s. $\partial\rho \neq \varnothing$. The convex conjugate $\hat{\rho}$ of $\rho$ is defined by $\hat{\rho}(\gamma) := \sup_{z \in \mathbf{R}^{n_2}} z\gamma - \rho(z)$ for all $\gamma \in \mathbf{R}^{n_2}$. $\hat{\rho}$ is known to be an extended normal convex integrand on $\mathbf{R}^{n_2}$ with $\tilde{P}$-a.s. $\partial\hat{\rho} \neq \varnothing$; see Theorem 5.3.13 in Molchanov (2005).

3.2. *Bounded quadratic growth.* Let $\rho$ denote a $\mathcal{P}$-measurable normal convex integrand on $\mathbf{R}^{n_2}$. We are going to need a quadratic growth bound on $\rho$, hence, we will assume throughout that $\tilde{P}$-a.s. $\rho \geq 0$, $\rho(0) = 0$ and

$$(3) \qquad \rho(z) \leq \frac{K}{2}|z|^2, \qquad z \in \mathbf{R}^{n_2},$$

for some constant $K > 0$. Note that then, for all $\gamma \in \mathbf{R}^{n_2}$,

$$\hat{\rho}(\gamma) := \sup_{z \in \mathbf{R}^{n_2}} z\gamma - \rho(z)$$
$$\geq \sup_{z \in \mathbf{R}^{n_2}} z\gamma - \frac{K}{2}|z|^2 = \frac{K^{-1}}{2}|\gamma|^2,$$

hence, we have $\tilde{P}$-a.s. $\hat{\rho} \geq 0$, $\hat{\rho}(0) = 0$ and a quadratic growth bound from below: $\hat{\rho}(z) \geq \frac{K^{-1}}{2}|z|^2, z \in \mathbf{R}^{n_2}$. Furthermore, we are going to need the following bounded growth condition on $\hat{\rho}$: For all $\gamma \in \mathbf{R}^{n_2}$, we assume $\tilde{P}$-a.s.

$$(4) \qquad \partial\hat{\rho}(\gamma) \subseteq B_{2K^{-1}|\gamma|},$$

where $B_r := \{z \in \mathbf{R}^{n_2} | |z| \leq r\}$. Note that then, for all $\gamma \in \mathbf{R}^{n_2}$, we have $\tilde{P}$-a.s.

$$\hat{\rho}(\gamma) = \int_0^1 \partial_\gamma^+ \hat{\rho}(s\gamma)\,ds$$
$$= \int_0^1 \sup_{z \in \partial\hat{\rho}(s\gamma)} z\gamma\,ds$$
$$\leq \int_0^1 2K^{-1}s|\gamma|^2\,ds$$
$$= K^{-1}|\gamma|^2 = \frac{(K/2)^{-1}}{2}|\gamma|^2.$$

Hence, we find similarly as above $\tilde{P}$-a.s. $\rho(z) \geq \frac{K/2}{2}|z|^2, z \in \mathbf{R}^{n_2}$. That is, under conditions (3) and (4), $\rho$ and $\hat{\rho}$ are both sandwiched between two

parabola. Alternatively, we could replace (3) by the stronger condition

(5) $$\partial \rho(z) \subseteq B_{K|z|}, \qquad z \in \mathbf{R}^{n_2}.$$

For $\rho, \hat{\rho}$ satisfying conditions (3) and (4), we can define $H^\rho(\cdot|Q^{\min})$ by equation (2) and we find immediately for all $Q \in \mathcal{M}$, $H^\rho(Q|Q^{\min}) < \infty$ iff $H(Q|Q^{\min}) < \infty$.

**4. A convex pricing functional.** In this section we are going to present a convex pricing functional, based on a generalized entropy penalty, which is compatible with prices for attainable contingent claims.

Assume condition (3) and consider the following convex quadratic BSDE:

(6) $$dY = -f(Z^1, Z^2)\, dt + Z^1\, dW^1 + Z^2\, dW^2, \qquad Y_T = \xi \in L^\infty,$$

with predictable random generator $f$ defined as $f(\omega, t, z^1, z^2) := -z^1 \lambda_t(\omega) + \rho(\omega, t, z^2)$ for $(\omega, t, z^1, z^2) \in \Omega \times [0, T] \times \mathbf{R}^{n_1} \times \mathbf{R}^{n_2}$, where we will often suppress the dependency on $(\omega, t)$ and write $f(Z^1, Z^2)$ or $f_t(Z^1, Z^2)$ instead.

BSDE (6) admits a unique solution $(Y^\xi, Z^1, Z^2) = (Y^\xi, Z^{\xi,1}, Z^{\xi,2})$ with square-integrable martingale part and $\sup_{t \in [0,T]} |Y_t^\xi| \in L^\infty$; see Kobylanski (2000). This allows us to define the following map $F: L^\infty \to \mathbf{R}$ by

(7) $$F(\xi) := F^\rho(\xi) := Y_0^\xi, \qquad \xi \in L^\infty.$$

It has been shown by Kobylanski (2000) that $F$ is continuous with respect to $\|\cdot\|_\infty$-norm and by the comparison principle, convexity of the generator $f$ implies $F$ to be convex.

Clearly, strong continuity is a desirable property of any pricing functional: Approximately, equal derivatives (with respect to $L_\infty$-norm) should have approximately the same prices. Admittedly, we do not have a *cogent* argument that all sensible pricing functionals necessarily should be convex. One could claim somewhat vaguely that convexity supports mitigation of risks by encouraging diversification and risk sharing, but in our opinion such arguments better apply if the pricing functional is in addition positively homogeneous. For the moment convexity is just a very useful technical property which we can not dispense with. However, in Section 5 we are going to see in which sense the risk immanent in $\xi$ enters the price $F(\xi)$, and this interpretation is very much based on the convexity of $F$, respectively $f$.

Since norm-continuity for convex functionals implies weak-$*$ lower semicontinuity by Mazur's lemma, we find $F$ to be convex and lower semicontinuous with respect to the $\sigma(L^\infty, L^1)$-topology on $L^\infty$ [see, e.g., Ekeland and Témam (1999)].

Consider the convex conjugate of $F$ with respect to the $\sigma(L^\infty, L^1)$-topology on $L^\infty$, restricted to $\mathcal{Q}$:

(8) $$\hat{F}(Q) := \hat{F}^\rho(Q) := \sup_{\xi \in L^\infty} E_Q[\xi] - F^\rho(\xi), \qquad Q \in \mathcal{Q}.$$

Let $\mathcal{H} = \mathcal{H}^T$ denote the set of $\mathbf{R}^{n_1}$-valued predictable processes $H$ on $[0, T]$ such that $E[\int_0^T |H_t|^2 \, dt] < \infty$ and for $V^H := (H\lambda) \cdot \mathbf{t} + H \cdot W^1$, $\sup_{0 \le t \le T} |V_t^H| \in L^\infty$. Note that for $H \in \mathcal{H}$ and $v_0 \in \mathbf{R}$, $(Y^\xi + v_0 + V^H, Z^1 + H, Z^2)$ solves BSDE (6) for terminal value $\xi + v_0 + V_T^H$. It follows that $F(\xi + v_0 + V_T^H) = F(\xi + v_0) = F(\xi) + v_0$ and we easily find $\hat{F}(Q) = \infty$ for $Q \notin \mathcal{M}$. In particular, $F$ is compatible with prices for attainable contingent claims since $F(v_0 + V_T^H) = v_0$.

THEOREM 4.1. *Under conditions* (3) *and* (4), *for all* $Q \in \mathcal{M}$, *we have*

(9) $$\hat{F}^\rho(Q) = H^\rho(Q|Q^{\min}).$$

*In particular,* $H^\rho(\cdot|Q^{\min})$ *is an extended weakly lower semi-continuous convex functional on* $\mathcal{M}$.

PROOF. Let $Q \in \mathcal{M}$ and denote by $q = \mathcal{E}(-\lambda \cdot W^1 + \gamma \cdot W^2)^T$ its density process. Set $\tau := \inf\{t \ge 0 | q_t = 0\} \wedge T$ and let $\xi \in L^\infty$ admit the BSDE representation $(Y^\xi, Z^1, Z^2)$. Note that for any sequence of stopping times $(T_n)_{n \ge 1}$, increasing to $\tau$, we have $E_Q[\xi] = \lim_{n \to \infty} E_Q[Y_{T_n}^\xi] = \lim_{n \to \infty} E[q_{T_n} Y_{T_n}^\xi]$ for $\xi \in L^\infty$, since $\sup_{t \in [0,T]} |Y_t^\xi| \in L^\infty$. Since for a local martingale $l$,

$$qY^\xi = Y_0^\xi + (q(Z^2\gamma - Z^1\lambda - f(Z^1, Z^2))) \cdot \mathbf{t} + l$$
$$= F(\xi) + (q(Z^2\gamma - \rho(Z^2))) \cdot \mathbf{t} + l,$$

we find

(10) $$E_Q[\xi] - F(\xi) = E\left[\int_0^T q_t(Z_t^2 \gamma_t - \rho_t(Z^2)) \, dt\right]$$

and

(11) $$\lim_{n \to \infty} E\left[\int_{T_n}^T q_t(Z_t^2 \gamma_t - \rho_t(Z^2)) \, dt\right] = 0.$$

Since $\gamma \cdot W^2$ is a local martingale on $[0, \tau)$, $|\gamma|^2 \cdot \mathbf{t}$ is on $[0, \tau)$ locally integrable. Under condition (4), and using the measurable selection theorem for random closed sets, we find a predictable $\mathbf{R}^{n_2}$-valued process $\tilde{Z}$ such that on $[0, \tau)$ $\tilde{P}$-a.s. $\tilde{Z} \in \partial \hat{\rho}(\gamma)$ holds. By condition (4), we have $|\tilde{Z}| \le 2K^{-1}|\gamma|$ on $[0, \tau)$. It follows that $|\tilde{Z}|^2 \cdot \mathbf{t}$ is locally integrable on $[0, \tau)$. Furthermore, since the process $-f(0, \tilde{Z}) \cdot \mathbf{t} + \tilde{Z} \cdot W^2$ is locally bounded on $[0, \tau)$, we can find a sequence of stopping times $(T_n)_{n \ge 1}$, $T_n < \tau$, increasing to $\tau$, and such that $\xi^n := -\int_0^{T_n} f_t(0, \tilde{Z}) \, dt + \int_0^{T_n} \tilde{Z}_t \, dW_t^2 \in L^\infty, n \ge 1$.

Since $\tilde{P}$-a.s. $\hat{\rho}(\gamma) = \hat{Z}\gamma - \rho(\hat{Z}) = \sup_{z \in \mathbf{R}^{n_2}} z\gamma - \rho(z)$ on $[0, \tau)$ iff $\tilde{P}$-a.s. $\hat{Z} \in \partial \hat{\rho}(\gamma)$ on $[0, \tau)$, and since for all $n \ge 1$, $\tilde{\xi}^n := \xi - Y_{T_n}^\xi \in L^\infty$, we find for $\hat{\xi}^n := \xi^n + \tilde{\xi}^n \in L^\infty$, $E_Q[\hat{\xi}^n] - F(\hat{\xi}^n) = E[\int_0^{T_n} q_t \hat{\rho}_t(\gamma) \, dt + \int_{T_n}^T q_t(Z_t^2 \gamma_t - \rho_t(Z^2)) \, dt]$. Hence,

$$\begin{aligned}
E_Q[\hat{\xi}^{n+1}] - F(\hat{\xi}^{n+1}) &\geq E_Q[\hat{\xi}^n] - F(\hat{\xi}^n) \\
&\geq E_Q[\xi] - F(\xi), \qquad n \geq 1,
\end{aligned} \tag{12}$$

and by Lemma 3.1, $\hat{F}(Q) = E[\int_0^T q_t \hat{\rho}_t(\gamma) \, dt] = H^\rho(Q|Q^{\min})$. □

Set $\tilde{F}(\xi) := \sup_{Q \in \mathcal{Q}} E_Q[\xi] - \hat{F}(Q) = \sup_{Q \in \mathcal{M}} E_Q[\xi] - \hat{F}(Q)$ for $\xi \in L^\infty$.

PROPOSITION 4.1. *Under conditions* (4) *and* (5), *for all* $\xi \in L^\infty$, $\tilde{F}(\xi) = F(\xi)$ *holds and there exists a probability measure* $Q^\xi \in \mathcal{M}$, $Q^\xi \sim P$ *with* $F(\xi) = E_{Q^\xi}[\xi] - \hat{F}(Q^\xi)$.

PROOF. We follow in part the proof of Theorem 2.1 in Frittelli (2000): Let $(Q^n)_{n \geq 1}$ be a sequence in $\mathcal{M}$ such that $E_{Q^n}[\xi] - \hat{F}(Q^n)$ increases to $\tilde{F}(\xi)$. Since $\{E_{Q^n}[\xi] | n \geq 1\}$ is bounded, $\{\hat{F}(Q^n) | n \geq 1\}$ is bounded too. By (3), we have $\tilde{P}$-a.s. $\hat{\rho}(z) \geq K^{-1} \frac{|z|^2}{2}, z \in \mathbf{R}^{n_2}$. Hence, $\hat{F}(Q) \geq K^{-1} H(Q|Q^{\min})$ for all $Q \in \mathcal{M}$ and $\{H(Q^n|Q^{\min}) | n \geq 1\}$ is bounded. It follows now from the Vallée–Poussin criterion [see, e.g., Dellacherie and Meyer (1982)] that $\{\frac{dQ^n}{dQ^{\min}} | n \geq 1\}$ is uniformly integrable with respect to $Q^{\min}$ and by the Dunford–Pettis compactness theorem, we can assume $(\frac{dQ^n}{dQ^{\min}})_{n \geq 1}$ to converge weakly in $L^1(Q^{\min})$ to $\frac{dQ^\xi}{dQ^{\min}}$ for a measure $Q^\xi \in \mathcal{Q}$. Since $S$ is locally bounded, we easily find $Q^\xi \in \mathcal{M}$. By weak convergence and weak lower semi-continuity of $\hat{F}$, we have $\hat{F}(Q^\xi) \leq \liminf_{n \to \infty} \hat{F}(Q^n)$, hence, $\tilde{F}(\xi) = E_{Q^\xi}[\xi] - \hat{F}(Q^\xi)$. Let $q = q^{Q^\xi}$ be given as $\mathcal{E}(-\lambda \cdot W^1 + \gamma \cdot W^2)$. $|\gamma|^2 \cdot \mathbf{t}$ is then locally integrable on $[0, \tau)$ for $\tau := \tau^{Q^\xi} \wedge T$ and by condition (4) $(q\hat{\rho}(\gamma)) \cdot \mathbf{t}$ as well. We have on $[0, \tau)$, for some local martingale $l$,

$$\begin{aligned}
qY^\xi - (q\hat{\rho}(\gamma)) \cdot \mathbf{t} - Y_0^\xi &= q \cdot Y^\xi + Y^\xi \cdot q + [q, Y^\xi] - (q\hat{\rho}(\gamma)) \cdot \mathbf{t} \\
&= (q(\gamma Z^2 - f(Z^1, Z^2) - Z^1 \lambda - \hat{\rho}(\gamma))) \cdot \mathbf{t} + l \\
&= (q(\gamma Z^2 - \rho(Z^2) - \hat{\rho}(\gamma))) \cdot \mathbf{t} + l.
\end{aligned}$$

Since $\tilde{P}$-a.s. $\gamma Z^2 - \rho(Z^2) \leq \hat{\rho}(\gamma)$ and $\gamma Z^2 - \rho(Z^2) = \hat{\rho}(\gamma)$ on $[0, \tau)$ iff $\tilde{P}$-a.s. $Z^2 \in \partial \hat{\rho}(\gamma)$ on $[0, \tau)$, we find by optimality of $Q^\xi$ that $\tilde{P}$-a.s. $\gamma Z^2 - \rho(Z^2) = \hat{\rho}(\gamma)$ on $[0, \tau)$. Hence, $qY^\xi - (q\hat{\rho}(\gamma)) \cdot \mathbf{t}$ is a local martingale on $[0, \tau)$ and the assertion follows from $F(\xi) = Y_0^\xi = \lim_{n \to \infty} E_{Q^\xi}[Y_{T_n}^\xi] - \hat{F}(Q^\xi) = \tilde{F}(\xi)$ for a localizing sequence $(T_n)_{n \geq 1}$, $T_n < \tau$, increasing to $\tau$. In order to show $Q^\xi \sim P$, observe that $\tilde{P}$-a.s. $Z^2 \in \partial \hat{\rho}(\gamma)$ on $[0, \tau)$ iff $\tilde{P}$-a.s. $\gamma \in \partial \rho(Z^2)$ on $[0, \tau)$, and condition (5) implies $\int_0^\tau |\gamma_t|^2 \, dt$ to be integrable. Hence $\gamma \cdot W^2$ is a square integrable martingale on $[0, \tau]$, implying $q_\tau > 0$ a.s. by the Doléans–Dade formula for the stochastic exponential and $\tau = T$. $\square$

Note that $F = \tilde{F}$ implies $F$ to be monotonic and, moreover, $\xi \geq 0$ and $P(\xi > 0) > 0$ implies $F(\xi) > 0$. Like strong continuity and weak lower semicontinuity, this property is desirable for any pricing functional. In the following section we give an interpretation of how the pricing functional penalizes for risk in terms of the instantaneous risk of an optimal tracking error.

Consider the predictable $\mathbf{R}^{n_2}$-valued process $\hat{\gamma} := \frac{Z^2 \rho(Z^2)}{|Z^2|^2}$, ($\frac{0}{0} = 0$). It is easy to check that for $\hat{q} := \mathcal{E}(-\lambda \cdot W^1 + \hat{\gamma} \cdot W^2)$, $\hat{q} Y^\xi$ is a local martingale on $[0, T]$. Similarly, as in Lemma 4.2 in Hu, Imkeller and Müller (2005), in order to show that $\hat{q}_T$ defines a probability measure $\hat{Q}^\xi \sim P$ in $\mathcal{M}$, it suffices by Theorem 2.3 in Kazamaki (1994) to show that $\hat{\gamma} \cdot W^2$ is a BMO martingale on $[0, T]$. Since $\tilde{P}$-a.s. $|\hat{\gamma}| \leq \frac{K}{2}|Z^2|$ by condition (3), it suffices to show that $Z^2 \cdot W^2$ or $M := Z^1 \cdot W^1 + Z^2 \cdot W^2$ is a BMO martingale on $[0, T]$.

LEMMA 4.1. $M^T$ is a BMO martingale.

PROOF. Set $E_\tau[\cdot] := E[\cdot|\mathcal{F}_\tau]$ for all stopping times $0 \leq \tau \leq T$. By translation invariance and uniqueness, we can without loss of generality assume $Y = Y^\xi \leq 0$. Using Itô's formula, we calculate

$$E_\tau[Y_T^2 - Y_\tau^2] \geq E_\tau\left[Y_T^2 - Y_\tau^2 - \int_\tau^T \left|\frac{Z_t^1}{\sqrt{2}} + \sqrt{2} Y_t \lambda_t\right|^2 dt\right]$$

$$= E_\tau\left[\int_\tau^T \frac{1}{2}|Z_t^1|^2 - 2Y_t^2 |\lambda_t|^2 + |Z_t^2|^2 - 2Y_t \rho_t(Z^2)\, dt\right]$$

$$\geq \frac{1}{2} E_\tau\left[\int_\tau^T |Z_t^1|^2 + |Z_t^2|^2\, dt\right] - E_\tau\left[\int_\tau^T 2Y_t^2 |\lambda_t|^2\, dt\right].$$

Since $\sup_{0 \leq t \leq T} |Y_t|, \sup_{0 \leq t \leq T} |\lambda_t| \in L^\infty$, we find $\sup_\tau E_\tau[\int_\tau^T |Z_t^1|^2 + |Z_t^2|^2\, dt]$ to be uniformly bounded. $\square$

Note that $Y^\xi$ is a uniformly bounded $\hat{Q}^\xi$-martingale. Hence, introducing a new asset with price process $Y^\xi$ to the market spanned by $S$ results again into an arbitrage-free market. However, we think of $\xi$ rather as a one-time OTC derivative deal that has to be priced. We do not expect a liquid market in the derivative $\xi$ with price process $Y^\xi$ to come into existence. (This is a quite realistic assumption: Even in deep option markets, only options near or at the money can really be regarded as liquid.) The seller is interested in hedging $\xi$ using the liquid assets $S$. In the following section we are going to see in which sense exactly the initial price $F(\xi)$ allows the seller to hedge $\xi$.

**5. Instantaneous risk.** We are going to work in the following setting: Let $\delta:\tilde{\Omega}\times\mathbf{R}^{n_1}\times\mathbf{R}^{n_2}\to\mathbf{R}$ be a normal convex integrand on $\mathbf{R}^{n_1}\times\mathbf{R}^{n_2}$. It is easy to check that then, for all $z^1\in\mathbf{R}^{n_1}$, the section $\delta_{z^1}:\tilde{\Omega}\times\mathbf{R}^{n_2}\to\mathbf{R}$, defined by $\delta_{z^1}(\tilde{\omega},z^2):=\delta(\tilde{\omega},z^1,z^2),(\tilde{\omega},z^2)\in\tilde{\Omega}\times\mathbf{R}^{n_2}$, is a normal convex integrand on $\mathbf{R}^{n_2}$. Furthermore, $\tilde{\omega}\mapsto\{(z^2,r)|\inf_{z^1\in\mathbf{R}^{n_1}}z^1\lambda(\tilde{\omega})+\delta(\tilde{\omega},-z^1,-z^2)\leq r\}=$ cl$\bigcup_{z^1\in\mathbf{Q}^{n_1}}\{(z^2,r)|z^1\lambda(\tilde{\omega})+\delta_{-z^1}(\tilde{\omega},-z^2)\leq r\}$ defines a closed random set. If, in addition, for $\delta_{z^2}:=\delta(\cdot,z^2)$, $\tilde{P}$-a.s. $\inf_{z^1\in\mathbf{R}^{n_1}}z^1\lambda+\delta_{z^2}(-z^1)=-\hat{\delta}_{z^2}(\lambda)>-\infty$, for example, if $\tilde{P}$-a.s. $\delta(z^1,z^2)\geq k(1+|z^1|^2)$ for some constant $k>0$, then $\rho:\tilde{\Omega}\times\mathbf{R}^{n_2}\to\mathbf{R}$, defined by $\rho(\tilde{\omega},z^2):=\inf_{z^1\in\mathbf{R}^{n_1}}z^1\lambda(\tilde{\omega})+\delta(\tilde{\omega},-z^1,-z^2)$, $(\tilde{\omega},z^2)\in\tilde{\Omega}\times\mathbf{R}^{n_2}$, is a normal integrand on $\mathbf{R}^{n_2}$. Assume that, for almost all $(\tilde{\omega},z^2)\in\tilde{\Omega}\times\mathbf{R}^{n_2}$, $z^1\lambda(\tilde{\omega})+\delta(\tilde{\omega},-z^1,-z^2)$ assumes its minimum for a unique $z^1\in\mathbf{R}^{n_1}$. It is easy to check that convexity of $\delta$ implies $\rho$ to be convex. Assuming now that $\delta(z^1,xz^2)$ increases in $x\geq 0$ for all $(z^1,z^2)\in\mathbf{R}^{n_1}\times\mathbf{R}^{n_2}$ and replacing $\delta$ by $\delta-\rho(0)$, we find for the corresponding $\rho$, $\rho\geq 0$ and $\rho(0)=0$. A uniform quadratic growth bound in the last variable of $\delta$ results into a uniform quadratic growth bound for $\rho$ as required in our setting.

Conversely, starting with $\rho$, consider $\delta$ defined by $\delta(z^1,z^2):=\rho(-z^2)+\frac{k^{-1}}{2}|\lambda|^2+\frac{k}{2}|z^1|^2$. It is then easy to check that the corresponding $\rho$ constructed as above equals the $\rho$ we have started with.

$\delta$ can be interpreted as an instantaneous risk measure: Let $(Y^\xi,Z^1,Z^2)$ be the solution to BSDE (6) and assume we are trying to track $Y^\xi$ by trading in $S$. Consider for $V^H:=Y_0^\xi+H\cdot S, H\in\mathcal{H}$, the tracking error increment [see Leitner (2006)]

$$d(V^H-Y^\xi)=(H\lambda+f(Z^1,Z^2))\,dt+(H-Z^1)\,dW^1-Z^2\,dW^2$$
$$=((H-Z^1)\lambda+\rho(Z^2))\,dt+(H-Z^1)\,dW^1-Z^2\,dW^2.$$

Assign to an increment $\mu\,dt+z^1\,dW^1+z^2\,dW^2$ the instantaneous risk $r(\mu,z^1,z^2):=-\mu+\delta(z^1,z^2)$ and note that this definition is an infinitesimal version of a convex risk measure, that is, it is translation invariant in the drift $\mu$ and convex in $(\mu,z^1,z^2)\in\mathbf{R}\times\mathbf{R}^{n_1}\times\mathbf{R}^{n_2}$. See Artzner et al. (1999) and Delbaen (2001) for coherent risk measures, and Föllmer and Schied (2002) for convex risk measures. For related results on dynamic convex risk measures, see Barrieu and El Karoui (2004, 2005, 2007) and Klöppel and Schweizer (2007).

DEFINITION 5.1. We say that $H\in\mathcal{H}$ has *instantaneously vanishing risk* for $\xi$ with respect to $\rho$ (or $\delta$) if the instantaneous risk $r((H-Z^1)\lambda+\rho(Z^2),H-Z^1,-Z^2)$ of the tracking error increment $d(V^H-Y^\xi)$ vanishes on $[0,T]$. $H$ has *nonnegative instantaneous risk* for $\xi$ if $r((H-Z^1)\lambda+\rho(Z^2),H-Z^1,-Z^2)\geq 0$ holds on $[0,T]$.

We have the following result:

PROPOSITION 5.1. *All $H \in \mathcal{H}$ have nonnegative instantaneous risk for $\xi$. $H$ has instantaneously vanishing risk with respect to $\rho$ iff (up to evanescence) $z^1 \lambda + \delta(-z^1, -Z^2)$ attains its minimum at $z_1 = Z^1 - H$.*

PROOF. We find for the instantaneous risk $r$ assigned to $d(V^H - Y^\xi)$

$$r = -((H - Z^1)\lambda + \rho(Z^2)) + \delta(H - Z^1, -Z^2)$$
$$= -\rho(Z^2) + (Z^1 - H)\lambda + \delta(-(Z^1 - H), -Z^2)$$
$$\geq -\rho(Z^2) + (Z^1 - \hat{H})\lambda + \delta(-(Z^1 - \hat{H}), -Z^2) = 0,$$

for $\hat{H} := Z^1 - h$, where $h$ satisfies $h\lambda + \delta(-h, -Z^2) = \inf_{z^1 \in \mathbf{R}^{n_1}} z^1 \lambda + \delta(-z^1, -Z^2)$. □

Note that it follows from the normality of $\delta$ that $h$ in the above proof can be chosen to be predictable. Under a bounded growth condition, the optimal $\hat{H} := Z^1 - h$ will be locally in $\mathcal{H}$. However, it is an open problem under which conditions $\hat{H} \in \mathcal{H}$ holds.

To give a simple example, consider $\delta(z^1, z^2) := c + \frac{k}{2}(|z^1|^2 + |z^2|^2)$ for predictable $c, k$, $k > 0$. For $c := \frac{k^{-1}}{2}|\lambda|^2$, it is easy to check that with $\rho$ constructed from $\delta$ as above, we have $\rho(z^2) = \frac{k}{2}|z^2|^2$ and $\hat{\rho}(z^2) = \frac{k^{-1}}{2}|z^2|^2, z^2 \in \mathbf{R}^{n_2}$. If $K^{-1} \leq k \leq K$ for some constant $K > 1$, we can apply our previous results, and we find $H^\rho(Q|Q^{\min}) = E_Q[\int_0^T \frac{k_t^{-1}}{2}|\gamma_t|^2 \, dt]$ for all $Q \in \mathcal{M}$, which can be interpreted as a *weighted* entropy functional. For the optimal tracking strategy $\hat{H}$, we find $\hat{H} = Z^1 + k^{-1}\lambda$ and $\hat{H} \cdot W^1$ to be a BMO martingale on $[0, T]$ by Lemma 4.1. This suggests that the right space of hedging strategies to work with could be the space of self-financing strategies such that the resulting value processes are BMO martingales with respect to $Q^{\min}$. However, solving quadratic BSDEs with unbounded terminal value seems to be quite difficult; see Briand and Hu (2006).

**6. Conclusions.** The advantage of our pricing method is that the dynamics of the tracking error provides an immediate feedback on the performance of the hedge. This is very convenient for P&L considerations practitioners are interested in. In contrast, having priced and sold a nonattainable contingent claim by expected (exponential) utility indifference, it is not clear how to hedge such a financial obligation in a good way from a (market) risk management point of view.

**Acknowledgments.** The author gratefully acknowledges a fruitful collaboration and continued support by Bank Austria through CDG. I thank the anonymous referees for helpful remarks.

Research Unit for Financial
 and Actuarial Mathematics
Institute for Mathematical Economics
Vienna University of Technology
Wiedner Hauptstrasse 8-10/105-1
A-1040 Vienna
Austria
E-mail: jleitner@fam.tuwien.ac.at